\documentclass[journal]{IEEEtran}
\usepackage{mathrsfs}
\usepackage{bbm}
\usepackage{amssymb}
\usepackage{bbding}
\usepackage{threeparttable}
\usepackage{colortbl}
\usepackage[mathcal]{euscript}

\usepackage{psfrag,calc,url,bm}

\usepackage{cite}

\usepackage{graphicx}

\usepackage{psfrag}

\usepackage{subfigure}
\usepackage{hyperref}
\usepackage{url}

\usepackage{stfloats}

\usepackage{amsmath}

\usepackage{float}

\usepackage{algorithmic}
\usepackage{diagbox}
\usepackage[ruled,linesnumbered,vlined]{algorithm2e}

\usepackage{color}

\usepackage{boxedminipage}

\usepackage{amsthm}

\usepackage{multirow}

\usepackage{setspace}

\usepackage{soul}
\usepackage{epstopdf}
\usepackage{booktabs}
\usepackage{svg}

\allowdisplaybreaks[3]

\begin{document}

\title{Graph Neural Network Enabled Pinching Antennas}

\author{Xinke Xie, Yang Lu,~\IEEEmembership{Member,~IEEE}, and Zhiguo Ding,~\IEEEmembership{Fellow,~IEEE}
\thanks{Xinke Xie and Yang Lu are with the School of Computer and Technology, Beijing Jiaotong University, Beijing 100044, China (e-mail: 21261022@bjtu.edu.cn,yanglu@bjtu.edu.cn).}
\thanks{Zhiguo Ding is with Department of Electrical Engineering and Computer Science, Khalifa University, Abu Dhabi 127788, UAE (e-mail: zhiguo.ding@ieee.org).}
}

\maketitle
\begin{abstract}

The pinching-antenna system is a novel flexible-antenna technology, which has the capabilities not only to combat large-scale path loss, but also to reconfigure the antenna array in a flexible manner. The key idea of pinching antennas is to apply small dielectric particles on a waveguide of arbitrary length, so that they can be positioned close to users to avoid significant large-scale path loss. This paper investigates the graph neural network (GNN) enabled transmit design for the joint optimization of antenna placement and power allocation in pinching-antenna systems. We formulate the downlink communication system equipped with pinching antennas as a bipartite graph, and propose a graph attention network (GAT) based model, termed bipartite GAT (BGAT), to solve an energy efficiency (EE) maximization problem. With the tailored readout processes, the BGAT guarantees a feasible solution, which also facilitates the unsupervised training. Numerical results demonstrate the effectiveness of pinching antennas in enhancing the system EE as well as the proposed BGAT in terms of optimality, scalability and computational efficiency.
\end{abstract}
\begin{IEEEkeywords}
Pinching Antennas, GNN, EE, BGAT.
\end{IEEEkeywords}

\section{Introduction}

\begin{figure}[t]
    \centering
\includegraphics[width=0.9\linewidth]{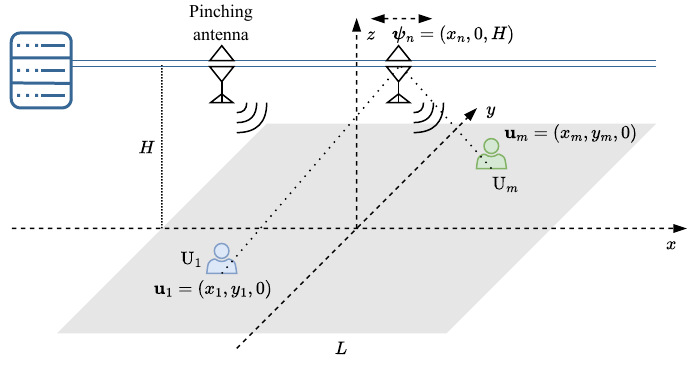}
    \caption{Illustration of the pinching-antenna system.}
    \label{sys_fig}
\end{figure}

Recently, the flexible-antenna systems, such as reconfigurable intelligent surfaces  \cite{ris} and fluid antennas \cite{fas}, have emerged as revolutionary techniques for reconfiguring wireless channel conditions. In particular, adjusting the phase shifts of signals or the deployment of antennas, the channel between the wireless transceiver which was once treated as a non-configurable parameter can be turned by a system designer. Nevertheless, most existing flexible-antenna systems are capable to provide small-scale  adjustments only, are capable to provide but they are not efficient to mitigate the large-scale path loss which dominates the channel conditions. To this end, the pinching antennas have been recently proposed as a novel evolution of  flexible-antenna systems \cite{pa0} as illustrated in \ref{sys_fig}, where the pinching antennas are deployed on a waveguide of arbitrary length, and activated by applying small dielectric particles on the waveguide. As a result, pinching antennas can be flexibly positioned close to the users to create strong line-of-sight communication and hence mitigate large-scale path loss\cite{pa1}. As demonstrated in \cite{pa2} and \cite{pa3},  by optimizing the deployment of the pinching antennas, the uplink transmission performance and the array gain can be significantly enhanced. However, the deeply coupled system parameters pose challenges to the development of efficient convex optimization (CVXopt)-based approaches for optimizing the pinching-antenna systems.  



On the other hand, deep learning (DL) has exhibited   great capabilities in handling the wireless optimization problems. In \cite{mlp,cnn,lugnn}, the multi-layer perceptrons (MLP), the convolutional neural network (CNN), and the graph neural network (GNN) have been employed to address the power allocation problems in wireless networks. These DL models achieve comparable performance to the CVXopt-based approaches but with a real-time inference speed. Among the DL models, the GNN yields superior  generalization and scalability performance thanks to its powerful capability of exploiting the graph topology of wireless networks \cite{gnnshen}. In \cite{ligat,dl2,dl3}, the GNN-based approaches were developed to achieve energy efficiency (EE) maximization, sum-rate maximization and max-min data rate, respectively, for multi-user multi-input-single-output (MISO) networks, and these GNN based approaches are scalable to the number of users. 




Applying the GNN to optimize pinching antennas presents new opportunities to realize the real-time and scalable transmit design. In this paper, we consider a joint placement and power allocation optimization problem subject to the transmit power budget and antenna spacing  in  a downlink communication system equipped with pinching antennas. Then, we formulate the considered system as a bipartite graph with user nodes and antenna nodes, and propose a graph attention network (GAT) based model, termed bipartite GAT (BGAT), to solve an EE maximization problem. With the tailored readout processes, the BGAT guarantees
feasible solutions, which also facilitates the unsupervised training. Numerical results demonstrate the effectiveness of  pinching antennas  in enhancing the system EE. Meanwhile, the BGAT outperforms the conventional design and DL benchmarks with the millisecond-level inference speed and scalability to the number of users.


\section{System Model and Problem Definition}

Consider a downlink time division multiple access (TDMA) based  communication system, as depicted in Fig. \ref{sys_fig}, where a base station (BS) equipped with $N$ pinching antennas activated on a single waveguide serves $M$ single-antenna users. The coordinate of the $n$-th ($n\in{\cal N}\triangleq\{1,2,\cdots,N\}$) pinching antenna is denoted by  ${{\bm\psi} }_n=[x_n,0,H]^T$, where $H$ is the height of the antenna. The $M$ users are uniformly distributed in a square with side length $L$, where the coordinate the $m$-th ($m\in{\cal M}\triangleq\{1,2,\cdots,M\}$) user is denoted by ${\bf u}_m =[x_m,y_m,0]$. 

Assume that the number of time slots in one TDMA frame is the same as that of users, and the $m$-th time slot is assigned to the  $m$-th user. The signal vector generated by the BS for the $m$-th user is denoted by 
\begin{flalign}
    &{\bf s}_m\left(\left\{p_n,{{\bm\psi} }_n\right\}\right) = \label{phase-shifted}\\
    &\left[ \sqrt{p_1}e^{-j\theta\left({{\bm\psi} }_1\right)},\sqrt{p_2} e^{-j\theta\left({{\bm\psi} }_2\right)}, \cdots, \sqrt{p_N} e^{-j\theta\left({{\bm\psi} }_N\right)} \right]^T s_m\in{\mathbb C}^N, \nonumber
\end{flalign}
where $p_n$ and $\theta({{\bm\psi} }_n)$ denote the power and the in-waveguide phase shift experienced at the $n$-th antenna, and $s_m$ denotes the signal passed onto the waveguide during the $m$-th time slot. Note that $\theta({{\bm\psi} }_n)$ is the phase shift for a signal traveling from the feed point of the waveguide to the $n$-th pinching antenna, and hence is a function of the coordinate of the $n$-th pinching antenna, i.e., $\theta({{\bm\psi} }_n) = 2\pi{ \| \boldsymbol{\psi}_0 - {{\bm\psi} }_n \|}/{\lambda}_{\rm g}$, where $ \boldsymbol{\psi}_0 $ denotes the coordinate of the feed point of the waveguide, and $\lambda_{\rm g} = c/(f_c n_{\rm neff})$ denotes the waveguide wavelength in a dielectric waveguide with $c$ , $f_c$ 
 and $n_{\rm neff}$ being the speed of light , the carrier frequency, and effective refractive index, respectively.

By treating the $N$ pinching antennas as conventional linear array antennas, the received signal at the $m$-th user is given by
\begin{flalign}
    &y_m =    {\bf h}^T_m\left( {\left\{ {{{\bm\psi} _n}} \right\}} \right){{\bf s}_m}\left(\left\{p_n,{{\bm\psi} }_n\right\}\right) + \omega_m,
    \label{rec_signal}
\end{flalign}
where $\omega_m$ denotes the additive white Gaussian noise with the power of $\sigma^2_m$, and ${\bf h}_m( {\{ {{{\bm\psi} _n}} \}} )$ denotes the channel vector to from $N$ pinching antennas to the $m$-th user , i.e.,
\begin{flalign}
    &{\bf h}_m\left( {\left\{ {{{\bm\psi} _n}} \right\}} \right) = \\
    &\left[ \frac{\sqrt{\eta}e^{-j\frac{2\pi\left \| {\bf u}_m-{ \boldsymbol{\psi} }_1\right \|}{\lambda}} }{\left \| {\bf u}_m-{ \boldsymbol{\psi} }_1 \right \| },  \cdots,  \frac{\sqrt{\eta}e^{-j\frac{2\pi\left \| {\bf u}_m-{ \boldsymbol{\psi} }_N\right \| }{\lambda}}}{\left \| {\bf u}_m-{ \boldsymbol{\psi} }_N \right \| }\right] ^{T},\nonumber
\end{flalign}
{where $\eta=c^2/(4\pi f_c)^2$ }.

Following \eqref{rec_signal}, the achieved information data rate  for receiving signal $s_m$ at the $k$-th user is given by
\begin{flalign}
&R_m\left(\left\{{p_{n}},{\bm \psi}_{n}\right\}\right) = \\
&\log\left(1+\frac{\left|\sum_{n\in{\cal N}}\sqrt{p_{n}}\frac{{\sqrt \eta  {e^{ { - j\left(\frac{{2\pi \left\| {{{\bm\psi} _m} - {\bm\psi} _{n}} \right\|}}{\lambda }+\theta_n\left({\bm\psi}_n\right)\right)} }}}}{{\left\| {{{\bm\psi}_m} - {\bm\psi} _{n}} \right\|}}\right|^2}{\sigma_m^2}\right).\nonumber
\end{flalign}


For the considered system, the EE maximization problem  is formulated as 
\begin{subequations}\label{po1}
\begin{align}
&\max_{\left\{p_n,{{\bm\psi} }_n\right\}}~\frac{\sum\nolimits_{m\in{\cal M}}~\tau R_m\left(\left\{p_n,{{\bm\psi} }_n\right\}\right)}{\sum\nolimits_{n\in{\cal N}} p_n+P_{\rm C}}\label{p1:a}\\
\text{s.t.}~& \sum\nolimits_{n\in{\cal N}} p_n \leq P_{\max},~p_n\ge0, \label{p1:b}\\
&{ \delta}_n\triangleq x_n - x_{n-1}  \ge \Delta,\label{p1:c}\\
&-D\le  x_n \le D,\label{p1:d}
\end{align}
\end{subequations}
where $\tau$ denotes the length of one time slot, $\Delta$ denotes the guard distance to avoid antenna coupling and $P_{\rm C}$ denotes the constant power consumption. Note that constraints \eqref{p1:c} and \eqref{p1:d} determine the adjustable areas of pinching antennas. Note that Problem \eqref{po1} is hard to address via traditional CVXopt-based  optimization techniques due to deeply coupled variables.

\section{Bipartite Graph Attention Network}




The considered system can be represented as a complete bipartite graph denoted by ${\cal G}=({\cal V},{\cal E})$, where $\cal V$ denotes the set of nodes with $|{\cal V}|=M+N$ and $\cal E$ denotes the set of undirected edges connecting nodes of different types. The $|{\cal V}|$ nodes are divided into two types, i.e., $M$ user nodes and $N$ antenna nodes. Specifically, the $m$-th user node has the feature of its location ${\bf u}_m$, while the $n$-th antenna node has the feature of the interval ${\delta}_n$ and its allocated power ${ p}_n$. The $\langle m,n\rangle$-th edge connecting the $m$-th user node and the $n$-th antenna node has the feature of  ${l}_{m,n}={ \| {\bf u}_m - {{\bm\psi} }_n \|}$. Note that ${\delta}_n$ and ${{\bm\psi} }_n$ require initialization manually.

\begin{figure}[t]
    \centering
\includegraphics[width=0.9\linewidth]{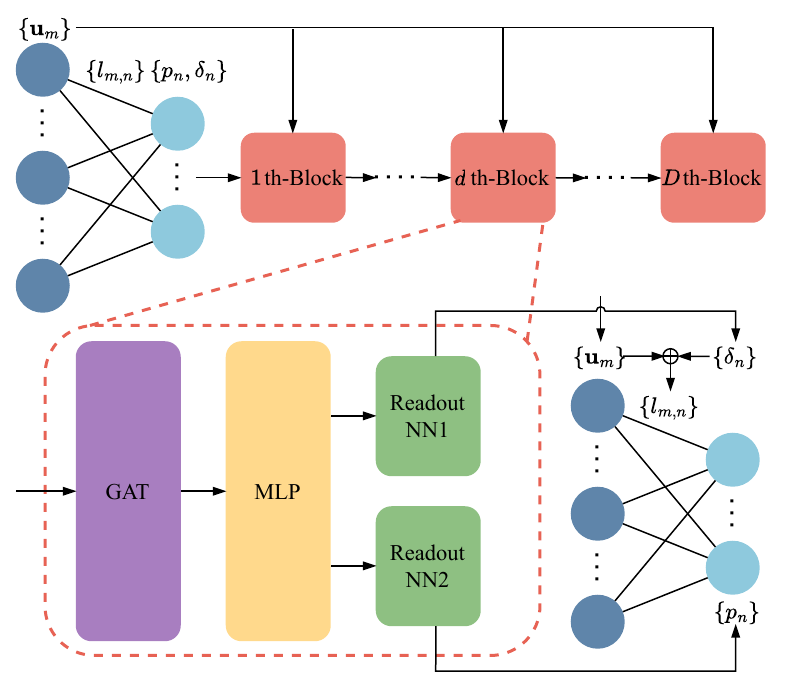}
    \caption{ Graph representation of the considered system and the architecture of the proposed model.}
    \label{model_fig}
\end{figure}

Following the ``learning to optimize" paradigm, we  construct and train a neural network with trainable parameters of ${\bm\theta}$ to realize the mapping\eqref{po1}, i.e.,
\begin{flalign}
    \left\{p_n,{{\bm\psi} }_n\right\} = \Pi_{\bm\theta}\left({\cal G}\right)
\end{flalign}
with $\{p_n,{{\bm\psi} }_n\}$ solving Problem $\rm P_1$ (near-) optimally. As the considered system is graph-structured, a particular form of GNN, termed BGAT, is utilized to build the mapping as illustrated in Fig. \ref{model_fig}, where the BGAT has $D$ blocks, each of which comprises three components, i.e., the GAT, the MLP, and two readout NNs. The detailed processes of the three components in one block are described as follows.


\subsection{GAT}
The GAT is to extract the graph-level feature of $\cal G$ to enhance the node features by exploring interactions among nodes. In the GAT block, the graph goes through four procedures including attention, aggregate, combination and update.

\subsubsection{Attention}

\begin{figure*}[b]
\hrule
\begin{flalign}\label{attention}
    &\alpha_{d,k,i,j} = \frac{
        \exp\left({\bf{a}}^{T}_{d,k}{\rm{LeakyReLU}}\left(
        {\bf{W}}_{{\rm s},d,k} {\bf{x}}_{d,i} + {\bf{W}}_{{\rm t},d,k} {\bf{x}}_{d,j} + {\bf{w}}_{{\rm e},d,k} {l}_{d,i,j}
        \right)\right)}
        {\sum_{v \in \mathcal{N}(i)}
        \exp\left({\bf{a}}^{T}_{d,k}{\rm{LeakyReLU}}\left(
        {\bf{W}}_{{\rm s},d,k} {\bf{x}}_{d,i} + {\bf{W}}_{{\rm t},d,k} {\bf{x}}_{d,v} + {\bf{w}}_{{\rm e},d,k} {l}_{d,i,v}
        \right)\right)}
\end{flalign}
\end{figure*}

Suppose that $K_d$ attention heads are employed in the GAT of the $d$-th ($ d\in{\cal D} \triangleq\{1,2\cdots,D\} $) block. The attention score associated with the $k$-th ($k\in{\cal K}_d \triangleq\{1,2\cdots,K_d\}$) attention head between the $i$-th node (user node or antenna node) and its $j$-th neighbor node is given by  \eqref{attention}, where  ${\bf{a}}\in{\mathbb{R}}^{\widetilde F_d}$ denotes the learnable attention vector, ${\bf{W}}_{{\rm s},d,k} \in {\mathbb{R}}^{{\widetilde F_d} \times F_d} $ and ${\bf{W}}_{{\rm t},d,k} \in {\mathbb{R}}^{{\widetilde F_d} \times F_d}$ denote the learnable linear transformation matrices for the source node feature and target node feature, respectively, ${\bf{w}}_{{\rm e},d,k}\in{\mathbb{R}}^{\widetilde F_d}$ denotes the learnable linear transformation vector for the edge feature, ${\widetilde F_d}$ and $F_d$ denote the corresponding dimensions, ${\bf x}_{d,i}$ and ${\bf x}_{d,j}$ denote the node features, $l_{d,i,j}$ denotes the edge feature, ${\mathcal N}(i)$ denotes the neighbor nodes of the $i$-th node, and ${\rm LeakyReLU}(\cdot)$ represents the LeakyReLU function.


\subsubsection{Aggregate}
With the obtained attention scores, the $i$-th node aggregates the weighted transformed features of its neighbor nodes to obtain its aggregated feature as 
\begin{flalign}
    {\bm \beta}_{d,k,i} = {\sum\nolimits_{j \in \mathcal{N}(i)}}\alpha_{d,k,i,j}{\bf{W}}_{{\rm t},d,k} {\bf{x}}_{d,j}.
\end{flalign}

\subsubsection{Combination}
The $K_d$ attention heads lead to $K_d$ aggregated features for the $i$-th node, i.e., $\{{\bm \beta}_{d,k,i}\}_k$. To fully utilized the aggregated features, we combine them for the $i$-th node. Besides, we add a residual connection to enhance the generalization ability of the model and alleviate the over-smoothing issue of GNN. The operation of combination is expressed as
\begin{flalign}
    {\widetilde{{\bf x}}}_{d,i} = {\rm Concat}_{k\in {\cal K}_d}\left( {\bm \beta}_{d,k,i} \right) + {\bf W}_{{\rm r},d,i}{\bf{x}}_{d,i},
\end{flalign}
where ${\bf W}_{{\rm r},d,m} \in {\mathbb R}^{({\widetilde F_d}\times K_d) \times F_d}$ denotes the learnable residual parameters, and ${\rm Concat}(\cdot)$ denotes concatenation operation.

\subsubsection{Update}
At last, ${\widetilde{{\bf x}}}_{d,i}$ is fed into an activation function to obtained an updated node feature as
\begin{flalign}
    {\bf x}_{d,i} \leftarrow {\rm ReLU}\left({\widetilde{{\bf x}}}_{d,i}\right),
\end{flalign}
where ${\rm ReLU}(\cdot)$ represents the ReLU function.


\subsection{MLP}

As defined, the original feature of one user node (e.g., ${\bf u}_m$) or antenna node (e.g., $\{{p}_n,\delta_n\}$) is two-dimensional. The MLP, stacking $T$ fully connected linear layers, is to transfer the node features to the required dimension. Denote $\overline {\bf x}_{d,i,t} \in {\mathbb R}^{{{{F}}_t}}$ by the input of the $t$-th ($t\in{\cal T}\triangleq\{1,2,...,T\}$) linear layer with ${{{{F}}_t}}$ being the dimension. Note that $\overline {\bf x}_{d,i,1}={\bf x}_{d,i}$ and ${{{F}}_{T+1}}=2$. The $t$-th linear layer is expressed as
\begin{flalign}
    \overline {\bf x}_{d,i,t+1} = {\rm ReLU}\left( {\bf W}_{d,t}\overline {\bf x}_{d,i,t} + {\bf b}_{d,t} \right),
\end{flalign}
where ${\bf W}_{d,t} \in {\mathbb R}^{{{F}}_t \times F_{t+1} }$ and ${\bf b}_{t}\in {\mathbb R}^{{{F}}_{t+1}}$ denote the learnable parameters and bias, respectively.  

\subsection{Readout NNs}

The readout NNs are responsible to yield feasible $\{p_n\}$ and $\{{\bm\psi}_n\}$ based on the obtained node features. We construct $\left\{\overline {\bf x}_{d,n,T+1} \in {\mathbb R}^2\right\}_{n\in {\cal N}}$ as a feature matrix denoted by ${\bf X}_d\in{\mathbb R}^{N \times 2 }$ with $[{\bf X}_d]_{(n,:)}=\overline {\bf x}_{d,n,T+1}$.


The first readout NN is to convert $[{\bf X}_d]_{(:,1)}$ to $\{{\bm\psi}_n\}$ satisfying \eqref{p1:c} and \eqref{p1:d}. First, the auxiliary vector is obtained as   
\begin{flalign}
   {{\bm \delta}_d = {\rm ReLU}\left({\rm MLP}_{d,\delta}\left(\left[{\bf X}_d\right]_{\left(:,1\right)}\right) \right)  \in {\mathbb R}^{N},}
\end{flalign}
where ${\rm MLP}_{d,\delta}(\cdot)$ denotes an MLP with learnable parameters ${\bf W}_{d,\delta}$, and ${\rm ReLU}(\cdot)$ ensures non-negative $[{\bm \delta}_d]_n$. Then, ${\bm \delta}_d$ is adjusted
\begin{flalign}\label{D_max}
   {\left[{\bm \delta}_{d+1}\right]_{n} \leftarrow \frac{B_{\max}}{{\rm max}\left( B_{\max}, \sum\nolimits_{i\in {\cal N}}\left[{\bm \delta}_{d}\right]_{i} \right)}\left[{\bm \delta}_{d}\right]_{n}}
\end{flalign}
 to satisfy $\sum_{i\in\cal N} [{\bm \delta}_{d+1}]_i \le B_{\max}\triangleq ( 2D-( N-1 )\Delta ).$ With ${\bm \delta}_{d+1}$, we obtain 
\begin{flalign}\label{Delta}
x_{d+1,n}=\left\{ \begin{array}{l}
\left[{{\bm \delta}_{d+1}}\right]_{n}-L,~n=1\\
 x_{d+1,n}+\left[{\bm \delta}_{d+1}\right]_{n}+\Delta-L,~n \neq 1 \end{array} \right.,
\end{flalign}
with $ {\bm \psi}_{d+1,n}\triangleq[x_{d+1,n}, 0, H]^{T}$, which satisfies (\ref{p1:c}) and (\ref{p1:d}). The feature of the $\langle m,n\rangle$-th edge is also upgraded by
\begin{flalign}\label{l}
    {l}_{d,m,n}={ \| {\bf u}_m - {{\bm\psi} }_{d+1,n} \|}.
\end{flalign}

The second readout NN is to convert $[{\bf X}_d]_{(:,2)}$ to $\{p_n\}$ satisfying \eqref{p1:b}. First, the power allocation can be obtained as follows:
\begin{flalign}
    {\bf p}_d = {\rm ReLU}\left( {\rm MLP}_{d,p}\left( \left[{\bf X}_d\right]_{\left(:,2\right)}\right)\right) \in {\mathbb R}^{N},
\end{flalign}
where ${\rm MLP}_{d,p}(\cdot)$ denotes an MLP with learnable parameters ${\bf W}_{d,p}$. Then, each element of ${\bf p}_d$ is fed into a scale function as 
\begin{flalign}
 \left[{\bf p}_{d+1}\right]_{n} \leftarrow \frac{P_{\max}}{{\rm max}\left( P_{\max}, \sum\nolimits_{i\in {\cal N}}\left[{\bf p}_{d}\right]_{i} \right)}\left[{\bf p}_{d}\right]_{n}.
\end{flalign}

At last, the output of the BGAT can be obtained by $p_n = \left[{\bf p}_{D+1}\right]_{n}$ and ${\bm \psi}_n = {\bm \psi}_{D+1,n}$.

\subsection{Loss function}
To guarantee a feasible solution, we adopt the unsupervised learning to train the proposed model. We directly use the objective function of Problem (\ref{po1}) to construct the loss function as
\begin{flalign}
    {\cal L} \left({\bm\theta}\right) = - \left.\frac{\sum\nolimits_{m\in{\cal M}}~R_m\left(\left\{p_n,{{\bm\psi} }_n\right\}\right)}{\sum\nolimits_{n\in{\cal N}} p_n+P_{\rm C}} \right|_{\bm\theta},
\end{flalign}
where 
\begin{flalign}
    {\bm\theta} \triangleq &\left\{ {\bf{a}}_{d,k}, {\bf{W}}_{{\rm s},d,k},{\bf{W}}_{{\rm t},d,k}, {\bf{w}}_{{\rm e},d,k}, \right.\\
    &\left.{\bf{W}}_{{\rm r},d,m}, {\bf W}_{d,t}, {\bf b}_{d,t}, {\bf W}_{d,\delta}, {\bf W}_{d,p} \right\},\nonumber
\end{flalign}
denotes all the learnable parameters for the proposed model, and it is observed that the dimensionality of ${\bm\theta}$ is independent of the number of users, which means the model can work in the user number variant scenarios even the problem size is unseen during the training phase.

\section{Numerical Results}

This section provides the numerical results to validate the effectiveness of the pinching antennas and the proposed BGAT. All the system parameters are shown in the table \ref{Simulation setting}.

\subsubsection{Datasets}
Each training set has $100,000$ samples while each test set has $1,000$ samples. In each sample, the $x$-coordinate and the $y$-coordinate of one user are sampled from a uniform distribution $U(-L,L)$, the inter-antenna interval ${ \delta}_n$ is initialized to $B_{\max}/(N-1)$, the edge feature $l_{m,n}$ is calculated by ${ \| {\bf u}_m - {{\bm\psi} }_n \|},$ and the power of one antenna is initialized by $P_{\max}/N$. 

\subsubsection{Initialization and training}
The learnable parameters are initialized according to He (Kaiming) method and the learning rate is initialized as $5\times10^{-5}$. The $Adam$ algorithm is adopted as the optimizer during the training phase. The batch size is set to $2,048$ for $1,000$ training epochs with early stopping to prevent overfitting. 
\subsubsection{Architecture of BGAT}
The BGAT under test has $5$ blocks with the same structure. The detailed structure of each block is showed in Table \ref{block}.
\begin{table}[t]
    \centering
        \caption{The Structure of each component in the proposed model}
    \begin{tabular}{c|c||c|c|c}
        \hline
         \multicolumn{2}{c||}{Component type}&In channel&Out channel&$\rm ReLU$  \\
         \hline
         \hline
         \multicolumn{2}{c||}{$\rm GAT$ }&$2$&${8\times4}^{\dagger}$&$\checkmark$\\
         \hline
         \multirow{2}{*}{$\rm MLP$}&Linear&$32$&$16$&$\checkmark$\\
         \cline{2-5}
         &Linear&$16$&$2$&$\checkmark$\\
         \hline
         \multirow{2}{*}{${\rm MLP}_{d,\delta}$}&Linear&$N$&$2\times N$&$\checkmark$\\
         \cline{2-5}
         &Linear&$2\times N$&$N$&$-$\\
         \hline
         \multirow{2}{*}{${\rm MLP}_{d,p}$}&Linear&$N$&$2\times N$&$\checkmark$\\
         \cline{2-5}
         &Linear&$2\times N$&$N$&$-$\\
         \hline

    \end{tabular}
    \begin{tablenotes}
    \small
    \item $\dagger$:The formae number means the out channel $\widetilde{F}_d $ actually, and the latter one is the number of heads $K_d$
    \end{tablenotes}
    \label{block}
\end{table}
\subsubsection{Baseline}
In order to evaluate the pinching-antenna system and the proposed BGAT numerically, the following three baselines are considered, i.e.,
\begin{itemize}
\item {Fixed-antenna scheme solved by CVXopt-based approach}: Detailed in Appendix A, termed ``Fixed". 
\item {Pinching-antenna system solved by MLP}:  A basic feed-forward  neural network, similar to \cite{mlp}, termed ``MLP". 
\item {Pinching-antenna system solved by GAT}:  Detailed in Appendix B, termed ``GAT". 
\end{itemize}

\begin{table}[t]
\centering
\caption{Simulation Parameters.}
\begin{tabular}{c|c }
\hline
{\bf Notation}               & {\bf Values}               \\ \hline
\hline
Number of Pinching antennas & {$N=\left\{4,8\right\}$} \\ \hline
 Number of users & {$M\in\{2,3,4,5\}$} \\ \hline
Size of range  & $L\in\{100\}$ \\ \hline
 Power budget   & $P_{\max}=30$ dBm \\ \hline
Carrier frequency & $f_c=6$ GHz \\ \hline
Height of antennas  & $H=5$ m \\ \hline
 Noise power  & $\sigma_{\rm B}^2=\sigma_{\rm E}^2=-90$ dBm \\ \hline
Minimum guide distance  & $\Delta=\lambda/2$ \\ \hline
Effective refractive index     & $n_{\rm neff}=1.4$ \\ \hline
Constant power consumption &$P_{\rm C} = 0.5$\\\hline
Coordinate of the feed point&${\bm\psi}_0 =[-D,0,d]$\\\hline
Length of one time slot & $\tau$ is normalized as $1$
\\\hline
\end{tabular}
\label{Simulation setting}
\end{table}

\subsubsection{Performance metrics}  The following evaluation metrics are considered on the test sets. 
\begin{itemize}
  \item 
  {Achievable EE}: The average EE (with the feasible solution) by the approach.
  \item 
  {Scalability}: The average achievable EE by the approach with problem sizes unseen in the training phase. 
  \item 
  {Inference time}: The average running time required to calculate the feasible solution by the approach. 
\end{itemize}

\subsection{Performance Evaluation}
The numerical results of the performance evaluation on test sets are provided in Table \ref{table:2}. The following is the analysis. 
\begin{table}[t]
\centering
\caption{Performance evaluation on test sets: EE [bit/Hz/J].}
\begin{tabular}{c|c|c||c|c|c|c}
\hline
$N$ &$M_{\rm Tr}$&$M_{\rm Te}$ &Fixed& MLP  & GAT & BGAT \\
 \hline
 \hline
 \rowcolor{blue!5} 
\cellcolor{white}~&\cellcolor{white}&\cellcolor{white}2&{28.17}&{26.35}&{30.59}&\bf{37.10}\\
  \rowcolor{orange!5}
 \cellcolor{white}~&\cellcolor{white}2&\cellcolor{white}3&\cellcolor{blue!5}{42.06}&{$\times$}&{46.71}&\bf{53.65}\\
   \rowcolor{orange!5} 
\cellcolor{white}\multirow{2}{*}{4}&\cellcolor{white}&\cellcolor{white}4&\cellcolor{blue!5}{56.07}&{$\times$}&{63.04}&\bf{70.71}\\
\cline{2-3}

~&~&3&\cellcolor{blue!5}{42.06}&\cellcolor{orange!5}{$\times$}&\cellcolor{orange!5}{51.71}&\cellcolor{orange!5}\bf{54.53}\\
 \rowcolor{blue!5} 
\cellcolor{white}~&\cellcolor{white}4&\cellcolor{white}4&{56.07}&{50.12}&{51.71}&\bf{71.95}\\
  \rowcolor{orange!5}
 \cellcolor{white}~&\cellcolor{white}~&\cellcolor{white}5&\cellcolor{blue!5}{70.45}&{$\times$}&{64.82}&\bf{89.37}\\
 \hline
 \multicolumn{3}{c||}{Inference time}&1.6 s& {\bf{2.5 ms}}&5.7 ms&8.7 ms\\
\hline
  \rowcolor{orange!5} 
\cellcolor{white}~&\cellcolor{white}~&\cellcolor{white}3&\cellcolor{blue!5}{39.47}&{$\times$}&{45.43}&\bf{53.91}\\
 \rowcolor{blue!5} 
\cellcolor{white}8&\cellcolor{white}4&\cellcolor{white}4&{52.76}&{52.85}&{61.55}&\bf{71.36}\\
  \rowcolor{orange!5}
 \cellcolor{white}~&\cellcolor{white}~&\cellcolor{white}5&\cellcolor{blue!5}{65.36}&{$\times$}&{77.77}&\bf{88.36}\\
 \hline
 \multicolumn{3}{c||}{Inference time}& 2.5 s& {\bf{2.7 ms}}&5.3 ms&8.5 ms\\
 \hline


\end{tabular}
\begin{tablenotes}
\footnotesize
\item $M_{\rm Tr}$/$M_{\rm Te}$ denotes the number of users in the training/test set.
\item $\times$ represents ``not applicable".
\end{tablenotes}
\label{table:2}
\end{table}

\subsubsection{Achievable EE (marked by blue-shaded areas)}
It is observed that the BGAT achieves higher EE compared with the fixed-antenna scheme, as the pinching antennas can effectively mitigate the path loss. Besides, it is also observed that the proposed BGAT surpasses the MLP and the GAT in all three cases. The increase in antennas and users significantly degrades the performance of the MLP and GAT, while having limited impacts on the performance of the BGAT.

\subsubsection{Scalability (marked by orange-shaded areas)}
It is observed that the proposed BGAT can also attain a satisfactory performance (with higher EE than the fixed-antenna scheme) in scenarios not encountered during the training phase. The BGAT is superior to GAT in terms of scalability performance, whereas MLP is ineffective.

\subsubsection{Inference time}
The inference time required by  well-trained DL models to obtain $\left\{p_n,{{\bm\psi} }_n\right\}$ for one realization of $\{{\bf u}_m\}$ is several milliseconds, which is significantly lower than the iterative CVXopt-based approach. Such a remarkable capability enables the DL models to be applicable under time-varying channel conditions. Besides, the inference time of DL models remains almost unchanged with the growth of system scale (determined by the numbers of $N$ and $M$), while that of the CVXopt-based approach increases exponentially.





\section{Conclusion}
This paper has investigated the GNN-enabled pinching-antenna systems by jointly optimizing the placement and power allocation of antennas. The bipartite graph representation for the considered system and the BGAT for addressing the considered problem were presented. Numerical results demonstrated the effectiveness of the pinching antennas in enhancing the system EE. Furthermore, the BGAT was shown to be applicable under  time-varying channel conditions and user number variant scenarios. 

\appendices
\section{Fixed-antenna scheme solved by CVXopt-based approach}

The fixed-antenna scheme deploys the antennas at $[0,0,0]$ with antenna spacing denoted by  $\Delta$. Define
\begin{flalign}
{\widetilde h}_{m,n}\triangleq\frac{{\sqrt \eta  {e^{ { - j\left(\frac{{2\pi \left\| {{{\bm\psi} _m} - \widetilde{\bm\psi} _{n}} \right\|}}{\lambda }+\theta_n\left(\widetilde{\bm\psi}_n\right)\right)} }}}}{{\left\| {{{\bm\psi}_m} - \widetilde{\bm\psi} _{n}} \right\|}}
\end{flalign}
with $\{\widetilde{\bm\psi} _{n}\}$ being the fixed locations of antennas. Then, Problems \eqref{po1} is  given by
\begin{subequations}\label{po2}
\begin{align}
&\max_{\eqref{p1:b}}~\frac{\sum\limits_{m\in{\cal M}}~\tau\log\left(1+\frac{\left|\widetilde{\bf h}_m^H {\bf p} \right|^2}{\sigma_m^2}\right)}{\sum\nolimits_{n\in{\cal N}} p_n+P_{\rm C}}
\end{align}
\end{subequations}
where $\widetilde{\bf h}_m\triangleq[{\widetilde h}_{m,1},\cdots,{\widetilde h}_{m,N}]^T$ and ${\bf p}\triangleq[\sqrt{p_1},\cdots,\sqrt{p_N}]^T$.

Define auxiliary variables 
\begin{flalign}
\left\{ \begin{array}{l}
\alpha_m=\log\left(1+\frac{\left|\widetilde{\bf h}_m^H {\bf p} \right|^2}{\sigma_m^2}\right),~\beta = \frac{\sum\limits_{m\in{\cal M}}~\log\left(1+\frac{\left|\widetilde{\bf h}_m^H {\bf p} \right|^2}{\sigma_m^2}\right)}{\left\|{\bf p}\right\|^2+P_{\rm C}}\\
a = \ln\left(\beta\right),~b = \ln\left(\left\|{\bf p}\right\|^2+P_{\rm C}\right)
\end{array} \right..
\end{flalign}
Problems \eqref{po1} can be solved by iteratively solving the following approximated optimization problem:
\begin{subequations}
\begin{align}
&\max_{\left\{{\bf p},\alpha_m,\beta,a,b\right\}}~\tau\beta\label{p2:a}\\
\text{s.t.}~& 2{{\rm Re}} \left\{ {{\widetilde{\bf{p}}}^H{{\bf{h}}}_m} {{\bf{h}}^H_m} {\bf{p}} \right\} - {\left| {{\bf{h}}_m^H} \widetilde{\bf{p}} \right|^2} \ge \left(2^{\alpha_m}-1\right){\sigma_m^2},\\
&\sum\nolimits_{m\in{\cal M}}{\alpha_m} \ge e^{(a+b)},\\
&e^{{\widetilde a}}\left(1+a - {\widetilde a} \right) \ge \beta,  \\
& e^{{\widetilde b}}\left(1+b - {\widetilde b} \right) \ge \left\|{\bf p}\right\|^2+P_{\rm C},\\
&\eqref{p1:b}.
\end{align}
\end{subequations}
in a successive convex approximation manner with $\{\widetilde{\bf p},\widetilde{a},\widetilde{b}\}$ being feasible to Problem \eqref{po1}.

\section{Pinching-antenna system solved by GAT}

\begin{figure}[t]
    \centering
\includegraphics[width=0.9\linewidth]{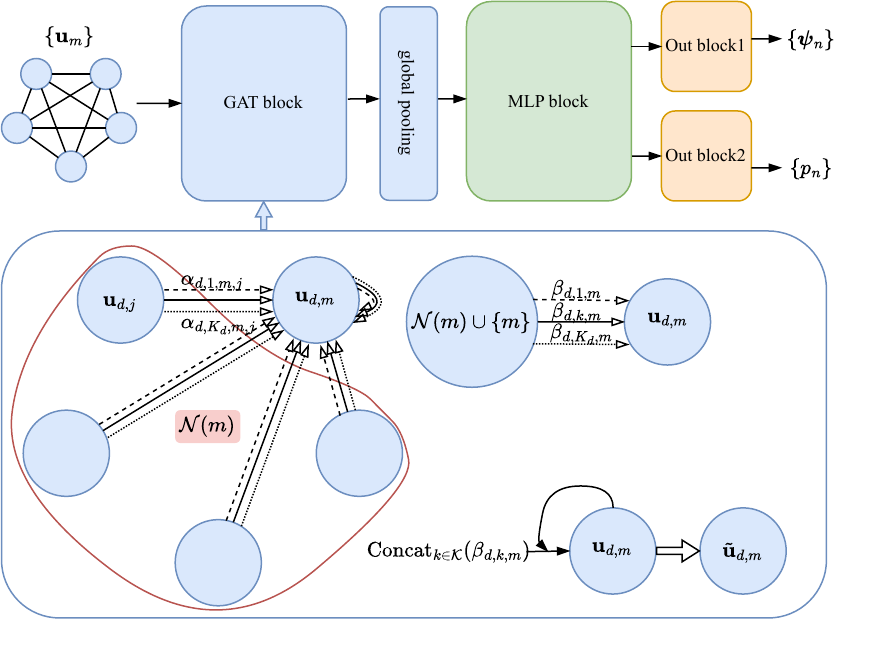}
    \caption{Illustration of the GAT.}
    \label{model_gat}
\end{figure}

The GAT for solving Problem \eqref{po1} is illustrated in Fig.\ref{model_gat}. The considered system is represented as a fully-connected graph denoted by ${\cal G^{\prime}}=({\cal V^{\prime}},{\cal E^{\prime}})$, where $\cal V^{\prime}$ denotes the set of nodes with $|{\cal V^{\prime}}|=M$ and $\cal E^{\prime}$ denotes the set of undirected non-feature edges. The $m$-th node has the feature of the location of the $m$-th user, i.e., ${\bm u}_{m}$. The GAT has four blocks, i.e., the GAT block comprising several GAT layers similar to Section III-A, the global pooling block to (e.g. max pooling) extract graph-level feature, the MLP block similar to Section III-B, and readout blocks similar to Section III-C.

\end{document}